\documentclass[twocolumn,showpacs,prb]{revtex4}

\usepackage{graphicx}
\usepackage{dcolumn}
\usepackage{bm}
\usepackage{color} 

\begin{document}

\title{
Iron-Based Heavy Quasiparticles in SrFe$_{4}$Sb$_{12}$: An Infrared Spectroscopic Study.
}

\author{Shin-ichi Kimura}
 \altaffiliation[Electronic address: ]{kimura@ims.ac.jp}
\affiliation{
UVSOR Facility, Institute for Molecular Science, Okazaki 444-8585, Japan
}
\affiliation{
School of Physical Sciences, The Graduate University for Advanced Studies (SOKENDAI), Okazaki 444-8585, Japan
}
\author{Takafumi Mizuno}
\author{Hojun Im}
\affiliation{
School of Physical Sciences, The Graduate University for Advanced Studies (SOKENDAI), Okazaki 444-8585, Japan
}
\author{Katsuyuki Hayashi}
\affiliation{
Department of Quantum Matter, ADSM, Hiroshima University, Higashi-Hiroshima 739-8530, Japan
}
\author{Eiichi Matsuoka}
 \altaffiliation{
Present address: Department of Physics, Tohoku University, Sendai, Japan
}
\affiliation{
Department of Quantum Matter, ADSM, Hiroshima University, Higashi-Hiroshima 739-8530, Japan
}
\author{Toshiro Takabatake}
\affiliation{
Department of Quantum Matter, ADSM, Hiroshima University, Higashi-Hiroshima 739-8530, Japan
}
\affiliation{
Institute for Advanced Materials Research, Hiroshima University, Higashi-Hiroshima 739-8530, Japan
}

\date{\today}

\begin{abstract} 
Temperature-dependent infrared reflectivity spectra of SrFe$_{4}$Sb$_{12}$ has been measured.
A renormalized Drude peak with a heavy effective mass and a pronounced pseudogap of 10~meV develops in the optical conductivity spectra at low temperatures.
As the temperature decreases below 100~K, the effective mass ($m^{*}$) rapidly increases, and the scattering rate ($1/\tau$) is quenched.
The temperature dependence of $m^{*}$ and $1/\tau$ indicates that the hybridization between the Fe $3d$ spins and the charge carriers plays an important role in determining the physical properties of SrFe$_{4}$Sb$_{12}$ at low temperatures.
This result is the clear evidence of the iron-based heavy quasiparticles.
\end{abstract}



\maketitle
\section{Introduction}
Recently, heavy quasiparticles or heavy fermions normally appearing in Ce- and Yb-based compounds have been observed in transition-metal compounds~\cite{Nidda}, for example LiV$_{2}$O$_{4}$~\cite{LiV2O4}, MnSi~\cite{MnSi} and ZrZn$_{2}$~\cite{ZrZn2,ZrZn2-dHvA} to name a few.
These materials have attracted attention because of their various physical properties, including those related to their performance as unconventional superconductors~\cite{Coleman}.
The origin of the various physical properties is believed to be related to the hybridization of charge carriers to localized spins.
Alkaline-earth-filled iron-antimony skutterudites ($A^{2+}$Fe$_{4}$Sb$_{12}$), including the SrFe$_{4}$Sb$_{12}$ system studied in this paper, are almost ferromagnetic systems with a paramagnetic Curie temperature ($T_{\rm C}$) of 53~K~\cite{Matsuoka, Matsumura, Schnelle}.
Alkali-filled iron-antimony skutterudites ($A^{+}$Fe$_{4}$Sb$_{12}$) possess an itinerant ferromagnetic character with a $T_{\rm C}$ = 80~K~\cite{ALJ1, ALJ2, Sheet}.
Both systems can be fundamentally explained by the self-consistent renormalization (SCR) theory~\cite{SCR}.
However, the Sommerfeld coefficient, $\gamma$, is enhanced by about 100 mJ/mol K$^{2}$ in $A^{2+}$Fe$_{4}$Sb$_{12}$~\cite{Matsuoka} compared to 53.2 mJ/mol K$^{2}$ predicted by a band structure calculation~\cite{Takegahara}.
In addition, the ratio between the enhanced coefficient $A$ of the quadratic electrical resistivity ($\rho$ = $AT^2$) and $\gamma$ is close to the Kadowaki-Woods value [1.0 $\times$ 10$^{-5} \mu \Omega$ cm K$^{-2}$/(mJ/mol K$^{2}$)$^{2}$]~\cite{Matsuoka}. 
This indicates that the hybridization between the charge carriers and the localized spins in $A^{2+}$Fe$_{4}$Sb$_{12}$ plays an important role at low temperatures.
The physical properties of SrFe$_{4}$Sb$_{12}$ are similar to those of the transition-metal compounds mentioned above.
Therefore, SrFe$_{4}$Sb$_{12}$ should exhibit similar physical properties to those of other transition-metal compounds as well as heavy fermion compounds.

The hybridization of the charge carriers to the localized magnetic moments forms heavy quasiparticles, which is called "Kondo effect".
In optical spectra, a renormalized Drude absorption due to these heavy quasiparticles is observed.
In heavy fermions and other transition-metal compounds mentioned above, the enhancement in the optical conductivity [$\sigma(\omega)$] at $\hbar \omega$~=~0~eV (coherent peak) and a peak at higher energy (incoherent peak) increase at low temperatures despite the fact that Drude curves for novel metals usually appear at high temperatures.
Such a change in $\sigma(\omega)$ has been observed in MnSi~\cite{MnSi-optics} and Sr$_{2}$RuO$_{4}$~\cite{Sr2RuO4} and in high-$T_{\rm C}$ cuprates~\cite{Puchkov-HTC, Dordevic}.
To clarify the appearance of the coherent and incoherent peaks in SrFe$_{4}$Sb$_{12}$ as well as the origin of the large $\gamma$, the temperature and magnetic field dependence of $\sigma(\omega)$ were measured and the effective mass ($m^{*}$) and scattering rate [$1/\tau$] were derived from the $\sigma(\omega)$ spectra.

$\sigma(\omega)$ in $A^{2+}$Fe$_{4}$Sb$_{12}$ ($A^{2+}$ = Ca, Ba) was previously reported to determine the origin of the pseudogap structure in divalent YbFe$_{4}$Sb$_{12}$~\cite{Joerg}.
Even though all of the $4f$ states are occupied in YbFe$_{4}$Sb$_{12}$, a pseudogap similar to the $c-f$ hybridization gap that generally appears in heavy fermion materials was observed above 10~meV~\cite{CeRu4Sb12}.
The paper {\it only} concluded that the origin of the pseudogap structure in YbFe$_{4}$Sb$_{12}$ was generated largely by the Fe $3d$ density of states.
In addition, $\sigma(\omega)$ for this system roughly reflects the unoccupied electronic states of the Fe$_{4}$Sb$_{12}$ frame~\cite{Takegahara, Kimura-PhysicaB}.
However, the previous paper did not clarify the origin of the heavy-quasiparticle-like character appearing in the thermodynamical properties~\cite{Matsuoka}.
In this communication, we point out the important role of the hybridization of carriers to the Fe $3d$ spins from the temperature and photon energy dependence of the effective mass and scattering rate of SrFe$_{4}$Sb$_{12}$ based on the temperature dependence of the $\sigma(\omega)$ spectrum {\it below} the pseudogap.

\section{Experimental}
A high-density polycrystalline SrFe$_{4}$Sb$_{12}$ sample was synthesized using a spark-plasma sintering technique previously reported~\cite{Matsuoka}.
The near-normal incident optical reflectivity spectra [$R(\omega)$] of SrFe$_{4}$Sb$_{12}$ were acquired from well polished samples by using 0.3~$\mu$m-grain-size Al$_{2}$O$_{3}$ wrapping film sheets.
Martin-Puplett and Michelson type rapid-scan Fourier spectrometers were used at photon energies ($\hbar \omega$) of 2.5~-~30~meV and 5~meV~-~1.5~eV, respectively, at sample temperatures between 7~-~300~K.
$R(\omega)$ under magnetic fields up to 6~T were also acquired by using a magneto-optical instrument at the beam line 6B of a synchrotron radiation ring, UVSOR-II, Institute for Molecular Science~\cite{IRMO}.
To obtain $R(\omega)$, the sample was evaporated {\it in-situ} with gold.
A reference spectrum was then measured.
To obtain $\sigma(\omega)$ via the Kramers-Kronig analysis (KKA), $R(\omega)$ at 300~K with zero magnetic field was measured over the energy range of 1.2 - 30~eV at the beam line 7B of UVSOR-II~\cite{BL7B}.
Since $R(\omega)$ above 1.2~eV does not significantly change, even though the temperature and magnetic field change, the $R(\omega)$ above 1.2~eV was connected to $R(\omega)$ under various conditions in the energy range below 1.5~eV.
In the energy ranges below 2,5~meV and above 30~eV, the spectra were extrapolated using the Hagen-Rubens function [$R(\omega)~=~1-(2\omega/\pi \sigma_{DC})^{1/2}$] and $R(\omega) \propto \omega^{-4}$, respectively~\cite{Wooten}.
After constructing $R(\omega)$ in the energy region from zero to infinity, KKA was performed to obtain $\sigma(\omega)$.

\begin{figure}[t]
\begin{center}
\includegraphics[width=0.38\textwidth]{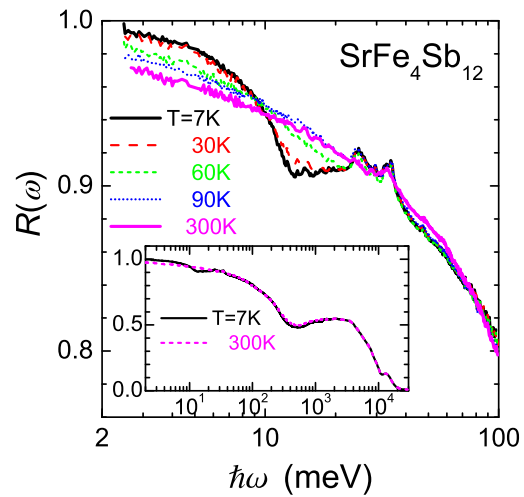}
\end{center}
\caption{
(Color online) The temperature dependent reflectivity spectrum [$R(\omega)$] in the photon energy range of 2 - 100~meV.
Inset: $R(\omega)$ at 7 and 300~K in the photon energy range of 2~meV - 30~eV.
}
\label{fig1}
\end{figure}
\section{Results and Discussion}

The temperature dependent $R(\omega)$ of SrFe$_{4}$Sb$_{12}$ obtained using the above method is shown in Fig.~\ref{fig1}.
$R(\omega)$ is a normal shape for a metal except for two significant peaks at 25 and 35~meV at the temperatures above 90~K.
The spectra also show a dip at around 12~meV that gets more pronounced with decreasing temperature.
Simultaneously, $R(\omega)$ below 10~meV increases to unity with decreasing temperature.
These spectral changes are similar to those of heavy fermion compounds~\cite{CePd3, Okamura}.
The two peaks at 25 and 35~meV are not discussed in this paper because the peaks originate from optical phonons~\cite{Joerg}.

\begin{figure}[t]
\begin{center}
\includegraphics[width=0.38\textwidth]{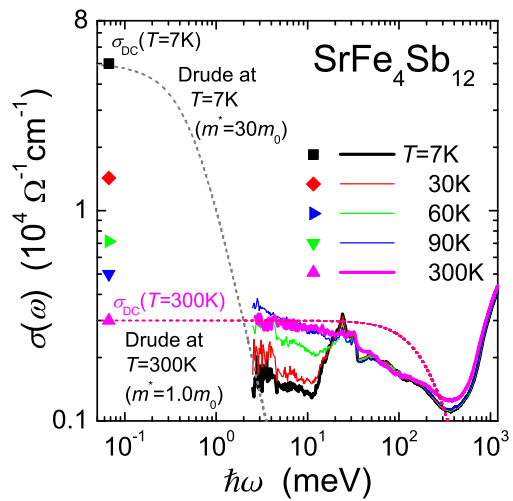}
\end{center}
\caption{
(Color online) The temperature dependence of the optical conductivity spectrum [$\sigma(\omega)$] of SrFe$_{4}$Sb$_{12}$ (solid lines) with the corresponding direct current conductivity ($\sigma_{DC}$, marks).
The classical Drude curves at 7 and 300~K using the $\sigma_{DC}$, the carrier density derived from the Hall coefficient and the expected effective mass ($m^{*}/m_0$~=~30 for $T$~=~7K and 1.0 for 300~K) are also shown.
}
\label{fig2}
\end{figure}

The temperature dependence of $\sigma(\omega)$ derived from $R(\omega)$ and the direct current conductivities ($\sigma_{DC}$) at the corresponding temperatures are shown in Fig.~\ref{fig2}.
At 300 and 90~K, the extrapolation of $\sigma(\omega)$ to 0~eV is consistent with the $\sigma_{DC}$s, which is consistent with the $\sigma(\omega)$ of a normal metal.
At 300~K, the Drude fitting curve with a single effective mass and single relaxation time (it is called the "{\it classical} Drude curve" hereafter) [$\sigma(\omega)~=~\sigma_{DC}/(1 + \omega^{2} \tau^{2})$] roughly represents the experimental $\sigma(\omega)$ below 200~meV.
Note that the upturn above 400~meV originates from the interband transition.
Here, $\sigma_{DC} = N_{eff} e^{2} \tau/m_{0}$, where $N_{eff}$ is the effective carrier density, $\tau$ the relaxation time and $m_{0}$ the electron rest mass.
$N_{eff}$ relates to $m^{*}$ with the function of $N_{eff} = N \cdot(m_{0}/m^{*})$.
The classical Drude curve at 300~K is calculated using the carrier density ($N$ = 6.2~$\times$~10$^{20}$ cm$^{-3}$) obtained from the Hall coefficient~\cite{Morimoto} and $m^{*}$ of 1.0~$m_0$ as evaluated by the {\it extended} Drude model analysis discussed later. 
With decreasing temperature, $\sigma(\omega)$ decreases at around 3~meV with a simultaneous increase in the $\sigma_{DC}$.
The contradictory temperature dependence in $\sigma(\omega)$ and $\sigma_{DC}$ indicates that $m^{*}$ increases and $1/\tau$ decreases with decreasing temperature.
It is noted that the classical Drude curve fitted at 7~K shown in Fig.~\ref{fig2} requires very heavy effective mass, for instance, $m^{*} = 30~m_{0}$.
However, the classical Drude curve at 7~K does not represent any of the obtained $\sigma(\omega)$.
This means that strong photon energy dependent $m^{*}$ and $1/\tau$ are expected.

\begin{figure}[t]
\begin{center}
\includegraphics[width=0.35\textwidth]{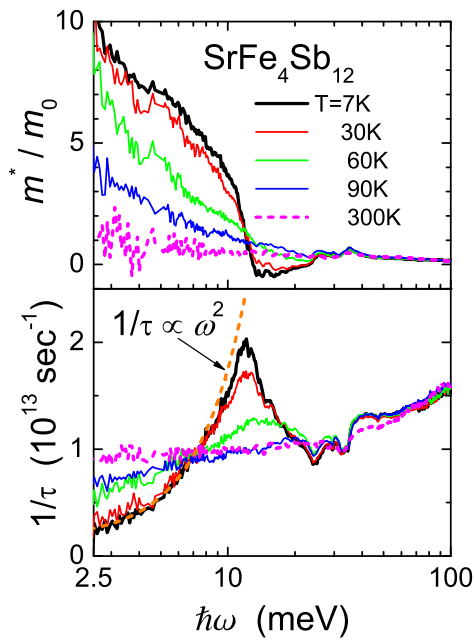}
\end{center}
\caption{
(Color online) The temperature dependent effective mass relative to the electron rest mass ($m^{*}/m_{0}$) and scattering rate ($1/\tau$) of SrFe$_{4}$Sb$_{12}$ as a function of photon energy.
A $1/\tau~\propto~\omega^{2}$ relation is also plotted for comparison, and holds for $\hbar \omega <$ 8~meV at temperatures below 30~K.
}
\label{fig3}
\end{figure}

To clarify the temperature and photon energy dependences of $m^{*}$ and $1/\tau$, we used the {\it extended} Drude model analysis using the real and imaginary parts of the dielectric function derived from the Kramers-Kronig analysis of $R(\omega)$~\cite{Kimura-RB6, Allen-NiSb, Puchkov-HTC}.
The $m^{*}/m_{0}$ and $1/\tau$ obtained from the followings;
\[
\frac{m^{*}}{m_{0}} = \frac{N e^2}{m_0 \omega} \cdot Im\left(\frac{1}{\tilde{\sigma}(\omega)}\right), 
\frac{1}{\tau} = \frac{N e^2}{m_0} \cdot Re\left(\frac{1}{\tilde{\sigma}(\omega)}\right).
\]
Here, $\tilde{\sigma}(\omega)$ is the complex optical conductivity derived from KKA of the corresponding reflectivity spectrum.
The $m^{*}/m_{0}$ and $1/\tau$ obtained from the analysis are plotted in Fig.~\ref{fig3}.
From the figure, both of $m^{*}$ and $1/\tau$ are almost constant at 300~K, with values of about 1.0~$m_{0}$ and 3.0~$\times$~10$^{13}$ sec$^{-1}$ below 20~meV, respectively.
This means that $\sigma(\omega)$ at 300~K can be described by the classical Drude model as shown in Fig.~\ref{fig2}.
With decreasing temperature, $m^{*}$ below 10~meV monotonically increases from (1.0~$\pm$~0.7)~$m_{0}$ at 300~K to (9.0~$\pm$~0.3)~$m_{0}$ at 7~K at $\hbar\omega$ = 3~meV.
The behavior of $1/\tau$, however, is not as linear, with, a peak at 12 meV increasing with decreasing temperature as well as a steeper decrease at lower energies, with a crossover point at about 6 meV.
At 7~K, $1/\tau$ is proportional to $\omega^2$ below 8~meV, indicating Fermi liquid behavior.
These results indicate that heavy quasiparticles are generated at low temperatures.

Similar temperature and photon energy dependences have been observed in heavy fermions~\cite{CePd3, Okamura} and high-$T_{\rm C}$ cuprates~\cite{Puchkov-HTC, Dordevic}.
In the case of SrFe$_{4}$Sb$_{12}$, $1/\tau$ has a peak at 12~meV, meaning carriers are scattered at that energy.
The peak energy is lower than those of heavy fermions (40~meV in YbAl$_{3}$~\cite{Okamura}, 17~meV in CePd$_{3}$~\cite{CePd3}) and of Sr$_{2}$RuO$_{4}$ (20~meV~\cite{Sr2RuO4}).
This indicates that the lower-energy excitation is related to the creation of the heavy quasiparticles.
An excitation energy of 12~meV corresponds to the energy gap of the "V"-shaped density of states originating from the hybridization band between the Fe~$3d$ and Sb~$5p$ states in the band structure calculation~\cite{Takegahara, Kimura-PhysicaB}.
Below the pseudogap, the hybridization between the charge carriers and the Fe $3d$ spins creates the heavy quasiparticles in SrFe$_{4}$Sb$_{12}$.

\begin{figure}[t]
\begin{center}
\includegraphics[width=0.38\textwidth]{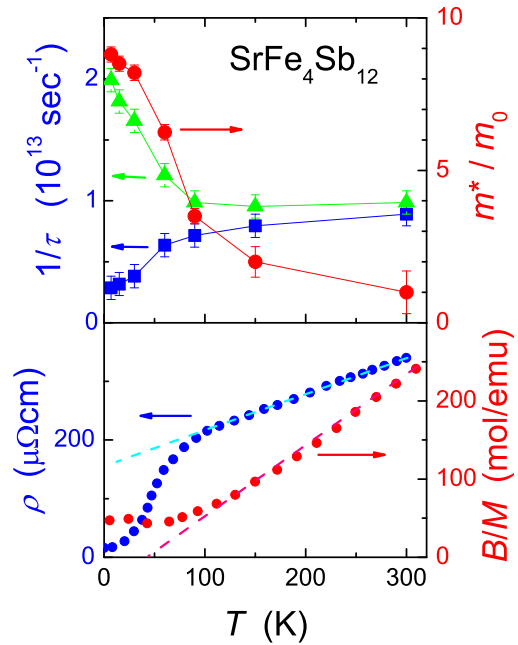}
\end{center}
\caption{
(Color online) $1/\tau$ at 3~meV (solid square) and 12~meV (solid triangle) and $m^{*}/m_{0}$ (solid circle) as a function of temperature.
The temperature dependence is compared with the electrical resistivity ($\rho$) and inverse magnetic susceptibility ($B/M$)~\cite{Matsuoka}.
}
\label{fig4}
\end{figure}

$m^{*}/m_{0}$, $1/\tau$ at 3.0~meV, and $1/\tau$ at the peak energy of the scattering rate (12~meV) are plotted in Fig.~\ref{fig4} as a function of photon energy.
The electrical resistivity ($\rho$) and the inverse magnetic susceptibility ($B/M$)~\cite{Matsuoka} are also plotted for comparison.
As can be seen in Fig.~\ref{fig4}, $m^{*}/m_{0}$ and $1/\tau$ dramatically change below 100~K.
In particular, $m^{*}$ at 7~K is about 9 times larger than that at 300~K.
This enhancement of the effective mass is the similar to the enhancement value of 8.7 of $\gamma$ in SrFe$_{4}$Sb$_{12}$ compared with SrRu$_{4}$Sb$_{12}$ (10 mJ/mol K$^{2}$)~\cite{Matsuoka-Os}.
The electronic structure near $E_{\rm F}$ of SrFe$_{4}$Sb$_{12}$ is similar to that of SrRu$_{4}$Sb$_{12}$ at 300~K because the $\sigma(\omega)$ of SrFe$_{4}$Sb$_{12}$ below 100~meV is the same as that of SrRu$_{4}$Sb$_{12}$ at 300~K~\cite{Kimura-full}.
Therefore, the mass enhancement at low temperatures has the same origin as the enhancement in $\gamma$.

The temperature dependences of $m^{*}/m_{0}$ and $1/\tau$, which indicate the anomaly at around 100~K are consistent with $\rho$ and $B/M$ as shown in Fig.~\ref{fig4}.
According to the nuclear quadrupole resonance (NQR) result, the ferromagnetic spin fluctuation is dominant above 100~K~\cite{Matsumura}.
Below 100~K, $(1/T_{1}T)_{spin}$ drops from the line of $(1/T_{1}T)_{spin}~\propto~\chi_{spin}$, which cannot be explained by the SCR theory.
This drop implies that the low-energy magnetic excitations associated with the spin fluctuations are almost inhibited at low temperatures.
The optical results indicate that the suppression of the spin fluctuations and the creation of the heavy quasiparticles observed in $\sigma(\omega)$ are coincident, or in the other words, the origin of these phenomena is the hybridization between the Fe $3d$ spins and the charge carriers.
This is consistent that the ratio between $A$ of $\rho$ = $AT^2$ and $\gamma$ in SrFe$_{4}$Sb$_{12}$ is close to the Kadowaki-Woods value as pointed out before~\cite{Matsuoka}.

\begin{figure}[t]
\begin{center}
\includegraphics[width=0.38\textwidth]{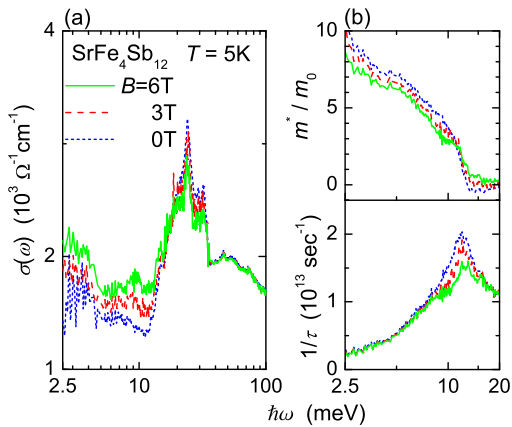}
\end{center}
\caption{
(Color online) (a) The magnetic field dependence of $\sigma(\omega)$ of SrFe$_{4}$Sb$_{12}$ and (b) the magnetic field dependent effective mass relative to the electron rest mass ($m^{*}/m_{0}$) and scattering rate ($1/\tau$) as a function of photon energy.
}
\label{fig5}
\end{figure}

If the mass enhancement that appears in $\sigma(\omega)$ originates from the hybridization of the charge carriers to the Fe $3d$ spins, the spectrum must change under magnetic fields.
Figure~\ref{fig5}(a) confirms this, as the peak in $\sigma(\omega)$ at around 20~meV decreases, and the dip below 12~meV disappears with increasing magnetic field strength.
$\sigma_{DC}$ also decreases with increasing magnetic field strength~\cite{Schnelle}.
The combination of the decreasing $\sigma_{DC}$ and the change in $\sigma(\omega)$ implies that the coherent and incoherent peaks collapse in the presence of a magnetic field.
The effective mass of the quasiparticles decreases with increasing magnetic field strength in spite that the scattering rate at the accessible lowest photon energy of 2.5~meV does not change as shown in Fig.~\ref{fig5}(b).
This is direct evidence of the creation of heavy quasiparticles due to the hybridization between charge carriers and the Fe $3d$ spins at low temperatures in SrFe$_{4}$Sb$_{12}$.
The magnetic field dependent effective mass is similar to that of a Ce-based heavy fermion skutterudite, CeRu$_{4}$Sb$_{12}$~\cite{Dordevic-B}.
The paper pointed out the effective mass as a function of magnetic field is suppressed as $1/B^{2}$.
However, since our accessible maximum magnetic field is 6~T, unfortunately we cannot follow the magnetic field dependence of the effective mass.
The optical measurement at higher magnetic fields for SrFe$_{4}$Sb$_{12}$ should be done for the complement.

In the above discussion, it is clear that the hybridization between the charge carriers and the Fe~$3d$ spins is important in understanding the physical properties in SrFe$_{4}$Sb$_{12}$.
However, no heavy quasiparticles appear in $A^{+}$Fe$_{4}$Sb$_{12}$ ($A^{+}$ = Na, K) because there is no coherent peak in $\sigma(\omega)$~\cite{Joerg-preprint}, in spite that $A^{+}$Fe$_{4}$Sb$_{12}$ possess an itinerant ferromagnetic character with a $T_{\rm C}$ = 80~K which is higher than that of $A^{2+}$Fe$_{4}$Sb$_{12}$.
On the contrary, trivalent LaFe$_{4}$Sb$_{12}$ is an enhanced paramagnetic metal.
The different physical character in $A$Fe$_{4}$Sb$_{12}$ originates in the difference in the carrier density due to the different valence number of guest atoms, {\it i.e.}, $A^{+}$Fe$_{4}$Sb$_{12}$ has the highest carrier density and LaFe$_{4}$Sb$_{12}$ has the lowest based on the positive Hall coefficient~\cite{Morimoto}.
The situation is similar to the carrier-density-controlled phase diagram in a heavy fermion system such as Ce(Ru$_{1-x}$Rh$_{x}$)$_{2}$Si$_{2}$~\cite{Miyako}.
In comparing $A$Fe$_{4}$Sb$_{12}$ with the phase diagram of Ce(Ru$_{1-x}$Rh$_{x}$)$_{2}$Si$_{2}$, $A^{+}$Fe$_{4}$Sb$_{12}$ and LaFe$_{4}$Sb$_{12}$ seem to be located in local and itinerant regimes compared to $A^{2+}$Fe$_{4}$Sb$_{12}$, respectively.
Therefore, the unconventional physical properties observed in $A^{+}$Fe$_{4}$Sb$_{12}$ are concluded to originate from the strong hybridization between the charge carriers and the Fe~$3d$ spins.

\section{Conclusion}
In conclusion, the temperature dependence of the optical conductivity of SrFe$_{4}$Sb$_{12}$ was measured in the photon energy range of 2.5~meV~-~30~eV.
With decreasing temperature, clear signatures of heavy quasiparticle behavior are found.
The optical effective mass is strongly enhanced below 10~meV and the scattering rate at 12~meV increases with decreasing temperature below 100~K.
The temperature dependence is consistent with the suppression of the spin fluctuations of the Fe $3d$.
This indicates that the hybridization between the Fe $3d$ spins and charge carriers plays an important role in the creation of the heavy quasiparticles in SrFe$_{4}$Sb$_{12}$ and also in other $A^{2+}$Fe$_{4}$Sb$_{12}$~\cite{Joerg} and YbFe$_{4}$Sb$_{12}$~\cite{CeRu4Sb12}.

\section*{ACKNOWLEDGMENTS}
We would like to thank Prof. K. Takegahara for fruitful discussion and Dr. Y. Sakurai for technical support for the reflectivity measurement in the VUV region.
This work was a joint studies program of the Institute for Molecular Science (2005) and was partially supported by a Grant-in-Aids; the COE Research (13CE2002), the priority area "Skutterudite" (No. 15072205) and Young Scientists (A) (No. 14702011) from MEXT of Japan.


\begin{thebibliography}{99}
\bibitem{Nidda} H.-A. Krug von Nidda, R. Bulla, N. B\"uttgen, M. Heinrich, and A. Loidl, Eur. Phys. J. B {\bf 34}, 399 (2003).
\bibitem{LiV2O4} S. Kondo, D. C. Johnston, C. A. Swenson, F. Borsa, A. V. Mahajan, L. L. Miller, T. Gu, A. I. Goldman, M. B. Maple, D. A. Gajewski, E. J. Freeman, N. R. Dilley, R. P. Dickey, J. Merrin, K. Kojima, G. M. Luke, Y. J. Uemura, O. Chmaissem, and J. D. Jorgensen, Phys. Rev. Lett. {\bf 78}, 3729 (1997).
\bibitem{MnSi} C. Pfleiderer, S. R. Julianand, and G. G. Lonzarich, Nature {\bf 414}, 427 (2001).
\bibitem{ZrZn2} C. Pfleiderer, M. Uhlarz, S. M. Hayden, R. Vollmer, H. v. L\"ohneysen, N. R. Bernhoeftand, and G. G. Lonzarich, Nature {\bf 412}, 58 (2001).
\bibitem{ZrZn2-dHvA} S. J. C. Yates, G. Santi, S. M. Hayden, P. J. Meeson, and S. B. Dugdale, Phys. Rev. Lett. {\bf 90}, 057003 (2003).
\bibitem{Coleman} P. Coleman and A.J. Schofield, Nature {\bf 433}, 226 (2005).
\bibitem{Matsuoka} E. Matsuoka, K. Hayashi, A. Ikeda, K. Tanaka, T. Takabatake, and M. Matsumura , J. Phys. Soc. Jpn. {\bf 74}, 1382 (2005).
\bibitem{Schnelle} W. Schnelle, A. Leithe-Jasper, M. Schmidt, H. Rosner, H. Borrmann, U. Burkhardt, J. A. Mydosh, and Y. Grin, Phys. Rev. B {\bf 72}, 020402(R) (2005).
\bibitem{Matsumura} M. Matsumura, G. Hyoudou, H. Kato, T. Nishioka, E. Matsuoka, H. Tou, T. Takabatake, and M. Sera, J. Phys. Soc. Jpn. {\bf 74}, 2205 (2005).
\bibitem{ALJ1} A. Leithe-Jasper, W. Schnelle, H. Rosner, N. Senthilkumaran, A. Rabis, M. Baenitz, A. Gippius, E. Morozova, J. A. Mydosh, and Y. Grin, Phys. Rev. Lett. {\bf 91}, 037208 (2003).
\bibitem{ALJ2} A. Leithe-Jasper, W. Schnelle, H. Rosner, M. Baenitz, A. Rabis, A. A. Gippius, E. N. Morozova, H. Borrmann, U. Burkhardt, R. Ramlau, U. Schwarz, J. A. Mydosh, Y. Grin, V. Ksenofontov and S. Reiman, Phys. Rev. B {\bf 70}, 214418 (2004).
\bibitem{Sheet} G. Sheet, H. Rosner, S. Wirth, A. Leithe-Jasper, W. Schnelle, U. Burkhardt, J. A. Mydosh, P. Raychaudhuri, and Yu. Grin, Phys. Rev. B {\bf 72}, 180407(R) (2005).
\bibitem{SCR} K. Ueda and T. Moriya, J. Phys. Soc. Jpn. {\bf 39}, 6687 (1975).
\bibitem{Takegahara} K. Takegahara and H. Harima, unpublished data.
\bibitem{MnSi-optics} F. P. Mena, D. van der Marel, A. Damascelli, M. F\"ath, A. A. Menovsky, and J. A. Mydosh, Phys. Rev. B {\bf 67}, 241101(R) (2003).
\bibitem{Sr2RuO4} T. Katsufuji, M. Kasai, and Y. Tokura, Phys. Rev. Lett. {\bf 76}, 126 (1996).
\bibitem{Puchkov-HTC} A.V. Puchkov, D. V. Basov, and T. Timusk, J. Phys.: Condens. Matter {\bf 8}, 10049 (1996).
\bibitem{Dordevic} S. V. Dordevic, C. C. Homes, J. J. Tu, T. Valla, M. Strongin, P. D. Johnson, G. D. Gu, and D. N. Basov, Phys. Rev. B {\bf 71}, 104529 (2005).
\bibitem{Joerg} J. Sichelschmidt, V. Voevodin, H. J. Im, S. Kimura, H. Rosner, A. Leithe-Jasper, W. Schnelle, U. Burkhardt, J. A. Mydosh, Yu. Grin, and F. Steglich, Phys. Rev. Lett {\bf 96}, 037406 (2006).
\bibitem{CeRu4Sb12} S. V. Dordevic, D. N. Basov, N. R. Dilley, E. D. Bauer, and M. B. Maple, Phys. Rev. Lett. {\bf 86}, 684 (2001).
\bibitem{Kimura-PhysicaB} S. Kimura, H. J. Im, Y. Sakurai, T. Mizuno, K. Takegahara, H. Harima, K. Hayashi, E. Matsuoka, and T. Takabatake, Physica B (2006) in press.
\bibitem{IRMO} S. Kimura, Jpn. J. Appl. Phys. {\bf 38} Suppl. 38-1, 392 (1999). 
\bibitem{BL7B} K. Fukui, H. Miura, H. Nakagawa, I. Shimoyama, K. Nakagawa, H. Okamura, T. Nanba, M. Hasumoto, and T. Kinoshita, Nucl. Instrum. Methods Phys. Res. A {\bf 467-468}, 601 (2001).
\bibitem{Wooten} F. Wooten, {\it Optical Properties of Solids} (Academic Press, New York, 1972).
\bibitem{CePd3} B. C. Webb, A. J. Sievers, and T. Mihalisin, Phys. Rev. Lett. {\bf 57}, 1951 (1986).
\bibitem{Okamura} H. Okamura, T. Michizawa, T. Nanba, and T. Ebihara, J. Phys. Soc. Jpn. {\bf 73}, 2045 (2004).
\bibitem{Morimoto} S. Morimoto, {\it et al.}, unpublished data.
\bibitem{Kimura-RB6} S. Kimura, T. Nanba, S. Kunii, and T. Kasuya, Phys. Rev. B {\bf 50}, 1406 (1994). 
\bibitem{Allen-NiSb} J.W. Allen and J.C. Mikkelsen, Phys. Rev. B {\bf 15}, 2952 (1977).
\bibitem{Matsuoka-Os} E. Matsuoka, S. Narazu, K. Hayashi, K. Umeo, and T. Takabatake, J. Phys. Soc. Jpn {\bf 75}, 014602 (2006).
\bibitem{Kimura-full} S. Kimura {\it et al.}, to be published.
\bibitem{Takahashi} Y. Takahashi, J. Phys.: Cond. Matter {\bf 13}, 6323 (2001).
\bibitem{Dordevic-B} S. V. Dordevic, K. S. D. Beach, N. Takeda, Y. J. Wang, M. B. Maple, and D. N. Basov, Phys. Rev. Lett. {\bf 96}, 017403 (2006).
\bibitem{Joerg-preprint} J. Sichelschmidt {\it et al.}, to be published.
\bibitem{Miyako} T. Taniguchi, Y. Tabata, and Y. Miyako, J. Phys. Soc. Jpn. {\bf 68}, 2026 (1999).

\end{thebibliography}
\end{document}